\documentclass[%
 reprint,
superscriptaddress,
 amsmath,amssymb,
 aps,
 pra,
]{revtex4-1}

\usepackage[utf8]{inputenc}
\usepackage{xcolor}
\usepackage{graphicx}
\usepackage{amsmath}
\usepackage{subfigure}
\usepackage{float}

\begin{document}

\title{Enhanced repetition codes for the cross-platform comparison of progress towards fault-tolerance}

\author{Milan Liepelt}
\affiliation{IBM Quantum, IBM Research -- Europe, Zurich}
\affiliation{Department of Physics and the Swiss Nanoscience Institute, University of Basel, Klingelbergstrasse 82, 4056 Basel, Switzerland}
\author{Tommaso Peduzzi}
\author{James R. Wootton}
\email{jwo@zurich.ibm.com}
\affiliation{IBM Quantum, IBM Research -- Europe, Zurich}

\date{\today}

\begin{abstract}

Achieving fault-tolerance will require a strong relationship between the hardware and the protocols used. Different approaches will therefore naturally have tailored proof-of-principle experiments to benchmark progress. Nevertheless, repetition codes have become a commonly used basis of experiments that allow cross-platform comparisons. Here we propose methods by which repetition code experiments can be expanded and improved, while retaining cross-platform compatibility. We also consider novel methods of analyzing the results, which offer more detailed insights than simple calculation of the logical error rate.

\end{abstract}

\maketitle

\section{Introduction and Motivation}

The last decade has seen a great deal of progress in the experimental demonstration of quantum error correction. A wide range of experiments have been carried out: from demonstrating detection of errors with the $[[2,0,2]]$ code~\cite{corcoles-15}, to $d=3$ and $d=5$ surface codes~\cite{krinner-22,zhao-22,acharya-23}; from extremely minimal demonstrations of twist defects~\cite{wootton-17} to full implementations~\cite{andersen-23,iqbal-23}. The scale of experiments can be expressed in terms of the number of qubits used, the code distance of the stored information and the number of syndrome measurement rounds. Thus far, published experiments with at least one repeated syndrome measurement have achieved codes with up to 49 qubits~\cite{acharya-23}, distances of up to 31~\cite{moses-23} and up to 50 syndrome measurement rounds~\cite{acharya-23}. Additionally, experiments with a single syndrome measurement round have used up to 97 qubits in a single code, and 280 qubits simultaneously over multiple codes~\cite{bluvstein-23}.

Though it is obvious that these experiments collectively demonstrate great progress, it can be hard to compare progress between different approaches. Surface codes on square lattice superconducting qubits~\cite{krinner-22,zhao-22,acharya-23} cannot easily be compared to Color codes with trapped ion qubits~\cite{ryan-anderson-21}, or to the codes implemented with heavy hexagonal superconducting qubits~\cite{chen-22,sundaresan-23}. This issue will be resolved once quantum error correction is able to support logical qubits, since the different approaches could naturally be benchmarked at the logical level. However, while we are still progressing towards fault-tolerance, it would be useful to have standardized methods to compare and contrast the capabilities of different forms of prototype quantum computing hardware, as we have for other aspects of quantum computing~\cite{amico-23}.

Our contribution is to propose such a standard methodology, and create the open-source software required to generate the required circuits and analyze the results. In this work we describe our approach as well as documenting its implementation in software.

Before diving in to the details, we first summarize the key points of our approach.

\begin{enumerate}
    \item We propose testing hardware by running a repetition code across the entire device. The flexibility of repetition codes, which can be easily adapted to different hardware constraints, should be used to full effect by moving beyond a standard linear layout.
    \item Since repetition codes cannot simultaneously detect bit and phase flip errors, we propose that $[[2,0,2]]$ codes are interleaved into the repetition code. This must be done such that each qubit is involved in a $[[2,0,2]]$ code at some point, but also such that the main repetition code maintains a high distance.
    \item Rather than simply calculating the logical error rate, internal details of the decoding process are analyzed. Specifically, we study the size of clusters formed by the Union Find decoder~\cite{delfosse-21}. We propose that the decay factor of the exponential decay of cluster size should be used as a device-wide measurement of error correction capability.
    \item The syndrome should also be analyzed to assess performance at the level of qubits and gates, using existing techniques for weight calculations of the decoding graph~\cite{wootton-20, wootton-22, wootton-thesis}.
\end{enumerate}

The rest of the paper is organized as follows. In Section \ref{sec:rep} we define the specific form of repetition codes that we consider. Then in Section \ref{sec:202} we propose an implementation of the $[[2,0,2]]$ code, and describe how they are interleaved. In Section \ref{sec:diagnostic} we define the quantities to be calculated from the results, to characterize performance at both the device-wide level, and at the level of individual qubits and gates. In Section \ref{sec:qiskit} we give an introduction to the implementation of these codes in Qiskit QEC~\cite{qiskitqec}, which allows for easy deployment of the experiments and analysis of the results. Then in Section \ref{sec:results} we provide example results from a 127 qubit IBM Quantum device, using 125 qubits for a distance 54 code with 10 syndrome measurement rounds.

\section{Generalized repetition codes} \label{sec:rep}

\subsection{Limitations of linear repetition codes}

The approach that we propose is inspired by the approach that already has widespread support: linear repetition codes. These have been implemented on  all major approaches towards quantum hardware~\cite{moussa-11,schindler-11,waldherr-14,wootton-20,takeda-22,moses-23}.

Linear repetition codes store a logical bit (rather than a logical \textit{qubit}) using product states of multiple qubits. The specific states chosen determine the kind of error that the code is affected by, with the standard encodings being for either bit-flip or phase-flip errors. The qubits of the code are regarded as being located along a line, with syndrome measurements that correspond to two-body parity checks on neighbouring pairs of qubits along this line.

Despite their prevalence, these codes have multiple limitations on their usefulness as a diagnostic of fault-tolerance. One is the restriction to a linear chain of syndrome measurements, which has the following drawbacks.
\begin{itemize}
\item Though a linear chain can typically be found that runs across any coupling map, many qubits may need to be left out.
\item The presence of faulty qubits or gates can break the line, preventing large codes from being probed.
\item Realizing effective fault-tolerance will require qubit connectivity that is better than that required for a linear repetition code. However, no advantages of these improved connectivities are seen in the diagnostic, other than the possibility of avoiding the previous point.
\end{itemize}

Another main limitation, which applies to any implementation of repetition codes, is the fact that they cannot detect a full set of errors. Typical encoding schemes allow either bit-flip or phase-flip errors to be detected, but not both simultaneously. Since the simultaneous detection of errors is a key requirement of fault-tolerance, this is a major discrepancy between this diagnostic and the processes that it hopes to inform.

With these limitations in mind, we propose a new standard for diagnostic repetition code experiments. These retain the features of repetition codes that make them useful as cross-platform benchmarks, while addressing the issues described above as much as possible.

\subsection{A new standard for repetition codes}

We will now lay out the details of the particular implementation of repetition codes that we consider, tailored to the task of serving as a diagnostic. In summary, our approach is to move beyond purely linear codes, and beyond codes with a purely bit- or phase-flip encoding. Instead, the layout of the code should be tailored to the device, and the encoding basis should alternate from qubit to qubit.

\subsubsection{Alternating encoding} \label{subsec:ae}

Typical implementations of repetition codes use either an encoding that protects against bit-flips or one that detects against phase-flips. For the former, the logical \texttt{0} state is encoded as $|0\rangle^{\otimes d}$, where $d$ is the number of repetitions, and logical \texttt{1} is encoded as $|1\rangle^{\otimes d}$. For the latter, \texttt{0} and \texttt{1} are encoded as $|+\rangle^{\otimes d}$ and $|-\rangle^{\otimes d}$, respectively. In the absence of errors, readout of the logical bit value can be done by measuring any of these qubits, using a $Z$-basis measurement in the former case and the $X$-basis for the latter.

These encodings are not typical of fully-functioning examples of quantum error correction, in which a full logical qubit can be stored. The most important difference is that neither is sensitive to a full set of errors: the former encoding cannot detect pure phase errors, such as an erroneously applied Pauli $Z$, and the latter cannot detect bit-flip errors, $X$.

Another difference is that the states used to encode the logical bit values are not entangled, but are instead product states. Indeed, by definition, all code qubits are simultaneously in an eigenstate of the same Pauli operator: $Z$ for the bit-flip encoding and $X$ for the phase-flip encoding. This is in contrast to stabilizer codes, for which logical information is encoded in highly entangled eigenstates of tensor products of Paulis that do not necessarily qubit-wise commute.

There is no easy fix to this issue. However, when using the repetition code simply as a diagnostic to measure the effects of noise in a system, there is one minimal action that should be taken: run both encodings. By getting information from both the bit-flip and phase-flip encoding, information about a full set of errors on each qubit is obtained, even if some does come from different runs with a different code. This tactic is used in many current experiments.

Another small improvement is to vary the basis used to encode from one qubit to the next. For example, for a $d=5$ linear repetition code, the logical bit values could be encoded as
\begin{eqnarray} \nonumber
\mathtt{0} &\rightarrow& |0\rangle \otimes |+\rangle \otimes |0\rangle \otimes |+\rangle \otimes |0\rangle \\ \nonumber
\mathtt{1} &\rightarrow& |1\rangle \otimes |-\rangle \otimes |1\rangle \otimes |-\rangle \otimes |1\rangle.
\end{eqnarray}
We refer to this as a $ZX$ encoding. Here, rather than using the $Z$-basis states to encode on all qubits as in the bit-flip encoding, or the $X$ basis states on all as in the phase-flip encoding, the basis alternates between the two. The complement to this is an $XZ$ encoding,
\begin{eqnarray} \nonumber
\mathtt{0} &\rightarrow& |+\rangle \otimes |0\rangle \otimes |+\rangle \otimes |0\rangle \otimes |+\rangle \\ \nonumber
\mathtt{1} &\rightarrow& |-\rangle \otimes |1\rangle \otimes |-\rangle \otimes |1\rangle \otimes |-\rangle.
\end{eqnarray}
In these, each qubit is only sensitive to bit-flip or phase-flip errors. However, around half of them are sensitive to one, and the other half to the other, and each region of the code has qubits sensitive to each. By running both variants, we again get a full set of information regarding noise.

These alternating encodings mimic fully-functioning examples of quantum error correction in that the code qubits are not all simply in an eigenstate of the same Pauli, and mean that at least isotropic multi-qubit errors will be detected at least somewhere. They are still a far cry from fully-functioning codes, and could not be used to protect a logical qubit. They are admittedly not a great improvement upon standard approaches to repetition codes, but they certainly do not provide a worse approach to diagnostics, and are are a step closer to mimicking the ideal behaviour. It is for this reason that we adopt the alternating encodings. They also inspire the name with which we refer to our standardization of repetition codes: `Alternating Repetition Codes' or `ARCs'.

\subsubsection{Syndrome measurements}

Repetition codes, like most other quantum error correcting codes, are typically defined using two sets of qubits. One set consists of the qubits that are explicitly used for the stored logical information, known as the code qubits or data qubits. The qubits referred to in the last subsection when explaining the encoding were exclusively of this form. We will use $d$ to refer to the number of code qubits, since for repetition codes this is also the code distance for the logical bit stored therein. The other set of qubits consists of those used as a resource during syndrome measurements, known as the auxiliary or ancillary qubits.

The syndrome measurements of repetition codes are two-body parity measurements, measuring an observable such as $Z \otimes Z$ or $X \otimes X$ on two code qubits. This measurement is mediated by an auxiliary qubit. Specifically, the auxiliary interacts pairwise with each code qubit and is subsequently measured to reveal the outcome of the syndrome measurement. The required qubit connectivity to implement such a measurement is therefore simply that the two code qubits must be connected to (i.e., able to interact pairwise with) their shared auxiliary. We refer to such a structure -- two code qubits connected to their shared auxiliary -- as a `link'.

Linear repetition codes are defined by placing the links of the code end-to-end. This either forms a line (for open boundary conditions) or a loop (for a periodic boundary). However, any arbitrary arrangement of links forms a valid repetition code. We will refer to this as a `link graph', where the code qubits and links form the nodes and edges of a graph, respectively. When a link graph forms a tree, then each syndrome measurement corresponds to an independent stabilizer generator. Otherwise, loops in the link graph correspond to a degree of redundancy. However, this does not lead to any complications, and so arbitrary link graphs can be considered.

\subsubsection{Scheduling}

Once the link graph has been specified, another important issue is the scheduling of entangling gates. For concreteness, let us now consider the bit-flip encoding of a repetition code, in which all links measure $Z \otimes Z$. The standard process for this is to first apply a CNOT gate between one of the code qubits and the auxiliary of the link, and then apply a CNOT between the other code qubit and the auxiliary. The auxiliary is then measured. The order in which the CNOT gates are applied is arbitrary, but they typically cannot be applied simultaneously.

For a given link graph, a schedule must be found for all the stabilizer measurements. One solution could be to measure the links one-by-one, applying the CNOTs in arbitrary order for each. One advantage of this is that the same auxiliary may be used for each measurement~\cite{moses-23}. A disadvantage is that the circuit depth required for a single round of syndrome measurement will then scale with the size of the code, which could lead to an unacceptable build up of noise.

Alternatively, a unique auxiliary may be used for each link, with a schedule that minimizes the circuit depth. For linear repetition codes, a syndrome measurement round can be performed with just two layers of CNOTs. One can apply, for example, CNOTs between each auxiliary and the code qubit to its left, and then do the same with the code qubit to the right. This then covers all required CNOTs. For repetition codes on more general link graphs, the number of layers is at least equal to the maximum degree within the graph.

The implementation of ARCs within Qiskit-QEC can automatically generate a schedule, but this will not neccessarily be optimal. The schedule may therefore need to be determined manually for best performance.

\subsubsection{Decoding} \label{subsec:d}

We will consider memory experiments, where logical information is initialized, a number of syndrome measurement rounds are applied, and all code qubits are then measured. Note that this final measurement is in the appropriate basis given the encoding of each qubit. The final information can be used to generate a final round of syndrome information by calculating the parity of the code qubit results for each link. It also provides the readout of the logical information: At least one qubit is designated for logical readout, with the result then providing an uncorrected outcome for the stored logical bit.

To obtain a corrected version of the logical information, all syndrome results must be taken into account. These give information about what errors have occurred, which can then be used to determine how the raw logical readout must be adjusted. The processing of the syndrome required for this is known as decoding.

The information regarding errors comes in the form of `error sensitive events'~\cite{chen-22}. For ARCs (and other stabilizer codes) these correspond to comparisons of two subsequent syndrome measurements. The same value will be repeated when no errors are present, so a difference in their outcomes signals that an error has occurred.

The above assumes that the auxiliary qubits are reset between rounds. The comparisons are complicated when this is not the case, since each outcome will be the \texttt{XOR} of all syndrome results for that link so far. In this case, no comparison is needed for the results of the second round. For all rounds thereafter, the comparison is made with the next-but-one round. For the results derived from final readout, the parity of these with the previous two rounds is used.

In either case, we can regard the error sensitive events as having a `time' and `space' coordinate. These are the round and link at which they were detected, respectively.

Measurement errors create a pair of such events, since the erroneous outcome features in two such comparisons. For an error on a code qubit, it will be detected on all adjacent links, and so the number of events that detect it is the number of such links.

We can also consider another form of error sensitive event: a difference between the raw logical readout and the correct value. This is not a quantity that we can compute, since knowledge of the correct value cannot be used in the decoding process. Rather, it is the job of decoding to look at the events detected in the syndrome, and use these to infer whether or not such events occurred for the logical readout.

Decoding algorithms often use the notion of a decoding graph. In this, the nodes of the graph correspond to the error sensitive events. For this reason, these events will henceforth be referred to as `nodes' in the context of decoding. Edges (or hyperedges in general) are defined by the action of a basic set of errors, typically single Pauli errors inserted at arbitrary points throughout the circuit. For every single such error we determine the set of nodes that detect it, and then add an edge or hyperedge to the graph that connects the corresponding nodes.

For linear ARCs, errors typically create pairs of nodes. This is due to each code qubit being incident upon two links (at least in the bulk). This correspondence between errors and pairs of nodes means that the decoding problem can be formulated as a minimum weight perfect matching problem, for which efficient and effective algorithms are known~\cite{higgott-23}.

For the case of any non-linear code, decoding must be done using another method. We consider HDRG-type decoders~\cite{hutter-15}, such as Bravyi-Haah~\cite{bravyi-13} and Union Find~\cite{delfosse-21}. These sift through different subsets of the nodes corresponding to the syndrome and determine whether each is neutral -- meaning that it could correspond to a valid set of errors -- and which logical events would be triggered in this case. By breaking down the full set of syndrome results into more manageable clusters of nodes, a final proposition for the net effect on the logical readout can be determined.

Determining whether a given cluster of nodes is neutral would be simple if only measurement errors occur. In this case, a valid set of errors simply implies that there should be an even number of nodes corresponding to each link. This means that any link with a odd number of nodes implies that errors must have occurred on code qubits. To determine the net effect of code qubit errors we first `flatten' the cluster, by finding the set of links for which there are an odd number of nodes. For this to correspond to a valid set of errors, the corresponding set of links must form an edge cut of the link graph. This means that it must be possible to bicolour the code qubits such that those that share a link within this set are of different colour while those sharing a link not within this set have the same colour. One such colour will then correspond to the code qubits on which errors occurred. Given this information, the potential effect of these errors on the logical readout can be determined.

The complexity of bicolouring all $d$ code qubits is $O(d)$. Analyzing sets of nodes in this way will therefore be computationally expensive, since considering even the smallest sets requires a complexity that increases with code size.

To improve the complexity, note that full bicolouring is not required. Only one of the two colour regions must be complete, since all other information can be easily deduced from this. For a set of nodes corresponding to a small number of errors, the computational effort can therefore be similarly small.

To achieve this, all queries for a given code are preceded by a one-time computational cost: analyzing the link graph to find a cycle basis. This information is then organized such that, for any given link, the basis cycles that include that link can be quickly found.

Then, for any query in which a given set of nodes is analyzed, the set of cycles in which these edges occur is then found. From this, the set of  code qubits around these cycles is found. This is a set of code qubits that it is necessary for the bicolouring to cover. Starting from one such code qubit, the bicolouring proceeds with a depth-first search until both: all necessary qubits are coloured, and; one colour region has stopped growing. The fact that the region is not growing, coupled with the fact that its boundary (as defined by the links in the given cluster of nodes) is fully coloured then shows that this region is complete.

\section{$[[ 2,0,2 ]]$ codes} \label{sec:202}

Scalable examples of codes that can detect a full set of errors typically have custom constraints on connectivity, meaning that no such code could provide a fair, cross-platform diagnostic. However, the smallest example of such a code has only very modest requirements, and is well-suited to cross platform implementation. This is the $[[2,0,2]]$ code~\cite{corcoles-15}, which has two code qubits and no stored logical information. Nevertheless, it has two syndrome measurements that together detect a full set of errors. For the standard (non-alternating) implementation of the code, these are $Z \otimes Z$ and $X \otimes X$, with the former detecting bit-flips, and the latter detecting phase-flips. Though we will wish to implement alternating versions of the code, we will nevertheless refer only to $Z \otimes Z$ and $X \otimes X$ measurements in the bulk of this subsection for simplicity.

A previous implementation of the $[[2,0,2]]$ code~\cite{corcoles-15} used four qubits in total: two as the code qubits and one auxiliary for each of the two measurements. For use with ARCs we instead introduce a time-ordered variant of the $[[2,0,2]]$ code, in which both syndrome measurements use the same auxiliary. This means that the measurements are done alternately. The previous implementation also initialized the code qubits in a known eigenstate of the syndrome measurements, so that comparing the outcomes of the measurements to these known values provided an error sensitive event. This was necessary because only one of each syndrome measurement was performed in that case. However, we will consider many rounds of syndrome measurement, allowing error sensitive events to be defined from subsequent outcomes of the same type of measurement. Specifically, bit-flip errors are detected when subsequent outcomes for $Z \otimes Z$ measurements differ, and phase-flips are detected when subsequent $X \otimes X$ outcomes differ. This is possible despite two $Z \otimes Z$ measurements having an $X \otimes X$ occur in between and vice-versa, because the two syndrome measurements commute.

For the codes we will consider, the standard measurements take an alternating form such as $Z \otimes X$. The $[[2,0,2]]$ code requires a pair of observables that anticommute qubit-wise, but commute overall. We use the simplest convention to obtain this, which is to reverse the tensor product. The roles of $Z \otimes Z$ and $X \otimes X$ in the above would therefore be played by, for the given example, $Z \otimes X$ and $X \otimes Z$, respectively.

\subsubsection{$[[ 2,0,2 ]]$ codes within ARCs}

Though repetition codes cannot simultaneously detect bit- and phase-flip errors across the entire code, this can be achieved on a link-by-link basis using $[[ 2,0,2 ]]$ codes. It is important to note that this will not contribute to the stability of the information stored in the code. Indeed it will detract from the stability, since several links of the repetition code need to be temporarily borrowed and repurposed to make the $[[ 2,0,2 ]]$ codes. However, recall that our aim is to benchmark circuits for quantum error correction, and characterize the errors therein. The use of $[[ 2,0,2 ]]$ codes helps towards this aim by allowing the simultaneous characterization of bit- and phase-flip errors, at least in a peicewise fashion.

In the following explanation of how $[[ 2,0,2 ]]$ codes can be inserted into ARCs we refer to the two observables measured for a given link are referred to as the `standard' and `conjugate' observables. The standard observable is the one that is normally measured on that link, while the conjugate observable is its reversed form.

The process of implementing a $[[ 2,0,2 ]]$ proceeds in a sequence of steps. During each step, all links but the given link and its neighbours (those which share a code qubit with the given link) are measured in the standard way. The steps are as follows.
\begin{enumerate}
\item Apply the standard syndrome measurements on all links, including the given link and its neighbours. A reset gate is then applied after all measurements, except for the auxiliaries of the neighbouring links which are left in the post-measurement state.
\item For the next seven rounds, measurements on the given link alternate between the conjugate measurement and the standard measurement. No measurement is applied to the neighbouring links.
\item Standard measurements are then applied on all links. For each of the two code qubits of the given link, one neighbouring link is chosen. The measurement outcome of this is used for a conditional Pauli. This Pauli is that from the conjugate observable on this code qubit. All auxiliaries are reset.
\item Standard measurements and resets are applied on all links.
\end{enumerate}

The conjugate measurements on the given link anticommute with the standard measurements on the neighbouring link. It is for this reason that the neighbouring links are only measured at the beginning and end of the process. Resets are not applied to the neighbouring links after the initial measurement so that the measurement result near the end of the process reflects the \texttt{XOR} of the results before and after the conjugate measurements. It therefore reflects whether or not the code qubits of the given link have been flipped (i.e., whether the conjugate observable has been applied to them). This information can therefore be used to undo the effect, which is the role played by the gates conditioned on the measurement outcomes. The inclusion of this operation allows the code to include a test of feedforward capabilities, which are required to some degree to achieve fault-tolerance, such as in the preparation of magic states~\cite{gupta-23}.

\begin{figure}
    \includegraphics[width=0.45\textwidth]{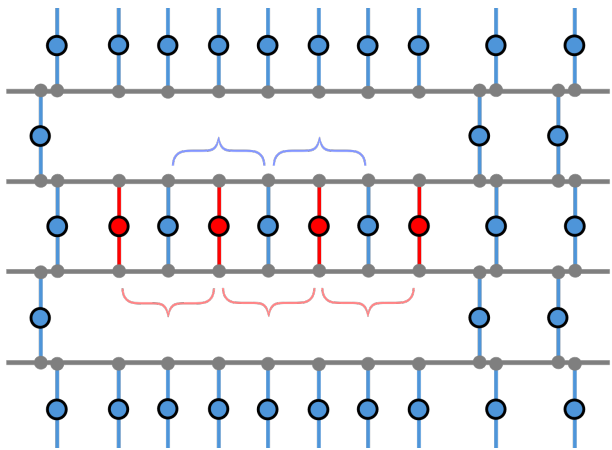}
    \caption{A representation of the circuit used for the $[[ 2,0,2 ]]$ process. As with standard circuit diagrams, time proceeds from left to right. A subset of four code qubits are shown, represented by the heavy grey lines. The central two code qubits are those for which the $[[ 2,0,2 ]]$ is applied. Standard syndrome measurements are represented as blue lines connecting code qubits, and conjugate measurements are represented by red lines. The measurements on auxiliary qubits are represented by coloured circles.}
    \label{fig:202circ}
\end{figure}

The most important error sensitive events in this process are those that can be used to simultaneously compare a full set of errors. When describing this we will refer to the standard errors (those detected by the standard measurements) and the conjugate errors (detected by the conjugate measurements). In Fig. \ref{fig:202circ}, these are shown by curly brackets

There are three events that detect the conjugate errors (depicted as red brackets). These are the three comparisons of subsequent conjugate measurements for the four conjugate measurements made. Since there is a time-like separation between these events, the errors detected by pairs of them will correspond to those on the auxiliary qubit. Conjugate errors on the code qubits are instead detected only by a single event, and hence by self-edges in the decoding graph. When analysing error rates corresponding to the edges (as will be explained in section \ref{subsec:m}), we will use the central self-edge here to measure the conjugate errors.

There are similarly events detecting standard errors during this period. Not including the initial and final standard measurements (during which the neighbouring links are also measured) there are three standard measurements and hence two error sensitive events (depicted as blue brackets). The self-edges corresponding to these will detect standard errors that occur at the same time as the conjugate errors described above. Analysis of these will therefore be used to compare standard errors to conjugate errors.

\section{Diagnostics derived from results} \label{sec:diagnostic}

So far we have described the form of the circuits to be run. In this section we will consider how the results of these can be processed to provide useful diagnostics of prototype quantum hardware. We will specifically look at two forms of diagnostic, which we refer to as microscopic and macroscopic. For the former we will analyze the syndrome to determine the error rates of individual qubits and gates within the quantum circuit. For the latter we will look at the behaviour of the code as a whole.

\subsection{Microscopic diagnostics} \label{subsec:m}

Syndrome information is designed to allow us to pinpoint when and where errors occur on a shot-by-shot basis. However, this also allows a powerful analysis of errors when we have many samples of syndrome information: we can determine the probabilities of different kinds of errors at different points within the circuit~\cite{wootton-22}. By doing so we can see exactly what noise is experienced by the qubits while implementing the code.

These microscopic error probabilities will be determined for specific qubits at specific times. We therefore require a standardized way to define a time coordinate. Rather than doing this based on the physical time during a circuit at which errors take place, we will use a more abstractly defined time coordinate based on the measurement rounds and the circuit depth. When defining times in this way, we use the convention that all gates take equal time. Time coordinates will be defined to have an integer part and a fractional part. The integer part will refer to the round during which the error occurs, and the fractional part is determined by the circuit depth within the syndrome extraction circuit.

To concretely define the fractional part we look to the schedule, which specifies when each of the entangling gates required in each round is applied by sorting them into a series of layers which can be applied simultaneously. Measurements are also inserted into the process to extract the required syndrome information. The current implementation of ARCs assumes that a single layer of simultaneous measurements applied at the end of the round. The total depth of a round, $\lambda$, is the sum of the number of entangling gate and measurement layers. The time coordinate of the point between the $j$th and $j+1$th layers within round $r$ is then defined as $r + j/\lambda$.

Errors correspond to hyperedges in the decoding graph in general, with the hyperedge for a given error corresponding to the set of nodes that detect it. However, since we will consider decoders that use edges and not hyperedges, and for simplicity and speed of analysis, we will restrict to graphs with simple edges. This means that, within the set of nodes that detect an error, we consider all possible pairs and add corresponding edges to the graph. For our purposes, errors therefore correspond to a pair of nodes. The only exception is the case of self-edges, corresponding to errors detected only by a single node.

Nodes have a time coordinate (the latter of the two rounds compared when the error is detected), and a space coordinate (the link on which they are detected). This information is used to determine a range of times at which the error could have occurred. When defining these ranges, we use the convention that the gates themselves are without error, with all errors occurring between the gates. The only exceptions to this are readout assignment errors, where result of a measurement is the opposite of the true result but with no erroneous effect on the qubit itself. For these, the time range will be that of the measurement gate.

The time range for given pairs of nodes is defined as follows.
\begin{itemize}
\item Two nodes that correspond to the same time but different links imply that the error occurred on the qubit shared between the two links between the two rounds whose values where compared. More concretely, the time at which the error could have occurred is bounded by the last time the shared qubit interacts with either of the auxiliaries for the links in the earlier round, and the first such interaction in the later round.
\item Two nodes that correspond to both different times and different links implies that the error must have occurred between the interactions with the two auxiliary qubits within the earlier round. 
\item Two nodes that correspond to the same link but for different times corresponds to a measurement error. This could be caused by an error on the auxiliary prior to measurement, or a falsely reported readout. When no resets are used, these can be distinguished by the two nodes being separated by one round or two rounds, respectively. The two cannot be distinguished when resets are used, so the time range covers the entirety of the first round for which the error was detected.
\end{itemize}

To determine the probabilities themselves we can use methods such as those of \cite{spitz-18}. However, in cases such as ours where the errors are more accurately described as being detected by hyperedges, such methods can lead to poor results~\cite{chen-22}. As an alternative we can use a method that we refer to as `naive', which performs a simple, approximate and yet robust analysis~\cite{wootton-20}. For any given edge of the decoding graph we look at all syndrome results, to determine the number of samples for which the syndrome contains neither of the edge's nodes, $n_{00}$, and the number for which it contains both of the nodes $n_{11}$. Since the errors that correspond to the given edge are the most likely to cause this difference, the two cases differ by no such error being present in the former case, and one such error being present in the latter. If we use $p$ to denote the probability of such an error, then to first order,
\begin{equation}
\frac{n_{11}}{n_{00}} \approx \frac{p}{1-p}.
\end{equation}
Each error probability can therefore be deduced by the corresponding ratio of $r = n_{11}/n_{00}$ by simply rearranging the above,
\begin{equation}
p \approx \frac{r}{1+r}.
\end{equation}
Note that $n_{00}$ and $n_{1}$ are proportional to the probability that the syndrome contains neither of the edge's nodes, $P_{00}$, and the probability that the syndrome contains both of the nodes, $P_{11}$, respectively. We could therefore equivalent use $r = P_{11}/P_{00}$

To get an idea of how approximate this analysis is, consider the case of an edge between two degree 10. This value was chosen because it is the maximum degree within the decoding graph of the specific version of the  code considered in Section \ref{sec:results}. In addition to $p$, the probability for the error corresponding to the given edge, we consider also $q$, the probability of error on the adjacent edges. The probabilities $P_{00}$ $P_{11}$, can then be easily calculated given the fact that a node appears when an odd number of the adjacent errors occur. 

\begin{figure}
    \subfigure{\includegraphics[width=0.45\textwidth]{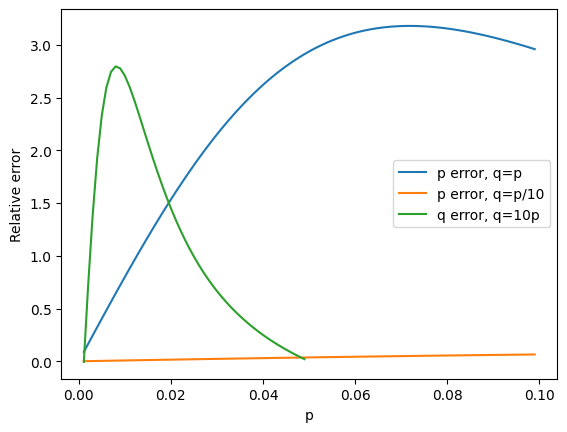}}
    \caption{A graph of the relative error of the inferred error probability for different values of $p$. In blue and orange are the relative errors with respect to $p$, for $q=p$ and $q=p/10$, respectively. In green is the relative error with respect to $q$ for $q=10 p$.}
      \label{fig:p_err}
\end{figure}

Relative errors for the inferred error probabilities are shown in Fig. \ref{fig:p_err}. The approximation is found to be good when $q \ll p$, and when $p=q$ is small. Otherwise, the relative error for the case of $p=q$ is found to peak at around $3$. So even when maximally incorrect, the approximation still holds to an order of magnitude. For $q \gg p$, however, the inferred value tends to more accurately reflect $q$ rather than $p$. In such cases, though the values we calculate will indeed reflect noise detected by the nodes considered, it does not reflect the intended noise source. This behaviour must therefore be kept in  mind when interpreting the microscopic values, particularly in cases where noise rates are high.

\subsection{Macroscopic diagnostics}


The most important diagnostic of a quantum error correcting code is the logical error rate. Since the whole purpose of the code is to preserve stored logical information, the probability that this fails determines how useful the code is. Determining this over the course of a computation is out of scope for current quantum hardware, and that of the near future. It is therefore instead common to consider the application of only a simple set of logical gates. The simplest example of all is to apply no logical gates: a so-called `memory experiment' where a state is initialized in a code, the syndrome measurements are then applied for a number of rounds $T$, and the state is finally read out. Comparing the decoded outcome to the expected outcome then gives the logical error rate.

A number of experiments have implemented such memory experiments in various codes. This includes repetition codes, where results for codes with up to $d=31$ have been published~\cite{moses-23}. However, since the logical error rate should decay exponentially with distance, the sampling complexity required to calculate these error rates is similarly exponential. Using these values as a diagnostic for ever increasing code sizes is therefore not scalable.

Furthermore, instead of single values for the logical error rate, it is typically more instructive to see how the value decays with increasing $d$ for a given $T$. One could also seek to define a lifetime for the logical information by increasing $T$ for a given $d$, until the logical error rate increases to a cutoff value. Such studies would provide a range of useful information about how changes in $d$ and $T$ affect the logical fidelity, and could be used to determine whether the expected exponential increasing lifetime is realized as $d$ increases. However, such studies have a significant overhead in the number of different values of $d$ and $T$ for which circuits must be run.

The repetition code is uniquely well-suited to addressing one aspect of the above problem. Simply by removing code qubits from a repetition code, we obtain a smaller and yet equally valid repetition code. We can therefore simply run the biggest code we can, and truncate the results to gain effective results for smaller codes~\cite{naveh-18}. However, the overhead in the number of syndrome measurement rounds still persists.

To address all of these issues, we will not analyze the logical error rate directly. Instead, we will extract more information from the decoder than simply what the final logical value is, in order to study the physical errors that cause the logical errors.

Specifically, we consider decoders that perform clustering on the syndrome: dividing it into subsets of the nodes which themselves form valid syndromes. Each of these clusters corresponds to a set of errors that corresponds to a bounded region around the cluster. This guess made by the decoder as to what errors caused the syndrome is then what is used to decode. In the case that the logical error rate is low -- and especially when it is too low to easily sample logical errors -- we can assume that the guesses made by the decoder are a good reflection of the errors. Since this diagnostic is designed for use under these conditions, we will henceforth assume that the clusters identified by the decoder do indeed correspond to regions in the code where large groups of errors occurred.

Sets of errors that correspond result in the same syndrome can be grouped into equivalence classes, depending on the correct decoding strategy that they require. Decoders then (ideally) determine which equivalence class is most likely for a given syndrome and decode accordingly. Logical errors happen when the set of errors that occurred does not belong to the most likely equivalence class. This typically corresponds to large groups of errors, located close enough that the syndrome cannot easily identify them individually. By analyzing the sizes of the clusters found by the decoders, and determining how likely it is to find groups of different sizes, we can then get a sense of how likely logical errors would be for codes of different sizes. For this we define the `size' of a cluster as the number of errors on code qubits required to create its flattened form, as described in Section \ref{subsec:d}.

For the repetition code and the clustering decoder we consider, we can expect the likelihood of a cluster of a given size to decay exponentially with the size. This is because, for any truncation of the code to one with $d$ code qubits (and hence distance), the probability of a logical error is the probability that there exists a cluster caused by at least $d/2$ code qubit errors~\cite{naveh-18,chen-21}. The expected exponential decay of logical error rate with distance therefore implies the exponential decay of cluster likelihood with size.

To characterize this decay, we apply the decoder and analyze the clusters to obtain a histogram of cluster sizes. We can fit a decay of the form $\rho^n$ to this, where $n$ is the size of the cluster.

In the limit of low noise, we would expect the distribution of errors to scale as $O\left(\left[ \frac{p}{\tau} \right]^n\right)$, where $p$ is an upper bound on error probabilities and $\tau$ is a code and decoder dependant quantity. We see from this that our fit parameter $\rho$ will have a linear dependence on $p$, and hence $\rho$ serves as a proxy for the noise level across the entire system.

Note that the $p/\tau$ here, and therefore $\rho$, serves a similar function to the $\Lambda$-factor of \cite{chen-21}. This would imply that $\tau$ corresponds to the noise threshold of the code. However we defer a further study of the relationship between $\rho$ and $\Lambda$ to future work.

Results for for repetition codes suited to current IBM Quantum hardware are shown in Subsection \ref{subsec:macro}, for which it is found that $\rho \approx 30 p$. The pseudothreshold of the code in this case is found to be around $5\%$. This agrees relatively well to the $\tau \approx 3.33 \%$ implied by the fit if we assume $\tau$ to be the threshold. The lack of full agreement could be due to the comparison of the pseudeothreshold to the full threshold, due to the entropic effects occurring away from the low noise limit, or simply because $\rho$ is not fully equivalent to $\Lambda$.

\section{ARCs in Qiskit-QEC} \label{sec:qiskit}

To provide a standardized form for diagnostics based on ARCs, we provide an implementation of them in software. This includes the means to create the circuits to be run on quantum hardware, as well as the methods required to analyze the results. This implementation is done in Python using the Qiskit package~\cite{qiskit}. It is included within Qiskit's framework for quantum error correction: Qiskit QEC~\cite{qiskitqec}.

\begin{figure}
    \includegraphics[width=0.25\textwidth]{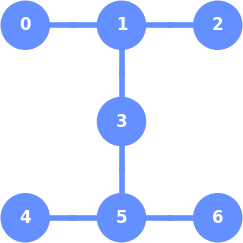}
    \caption{Layout of \texttt{ibm\_lagos}, showing the numbering used for the qubits.}
    \label{fig:lagos}
\end{figure}

In Qiskit QEC, circuits for codes are handled by corresponding \texttt{CodeCircuit} classes. For ARCs, the \texttt{ArcCircuit} class is used. This is imported and initialized as follows.
\begin{verbatim}
from qiskit_qec.circuits import ArcCircuit
code = ArcCircuit(links, T)
\end{verbatim}
Here \texttt{T} is the number of syndrome measurement rounds to be performed and \texttt{links} is a list of tuples that specify the links used. For example, given the coupling map of Fig. \ref{fig:lagos}, a $d=3$ code could be defined using qubits $0$, $3$ and $6$ as code qubits and $1$ and $5$ as auxiliaries. This corresponds to two links, which are expressed as follows.
\begin{verbatim}
links  = [(0,1,3), (3,5,6)]
\end{verbatim}

There are also a number of optional keyword arguments that can be specified when initializing \texttt{ArcCircuit}. The \texttt{schedule} argument can be used to specify a schedule for the entangling gates during each syndrome measurement round. For the $d=3$ example, such a schedule could be expressed as follows.
\begin{verbatim}
schedule = [[(0,1), (3,5)], [(3,1), (6,5)]]
\end{verbatim}
Here two layers of entangling gates are used. In the first, a gate is applied to qubits $0$ and $1$ at the same time as one on qubits $3$ and $5$. Then a gate is applied to qubits $3$ and $1$ at the same time as one on qubits $6$ and $5$. Note that the code qubit is listed first and the auxiliary listed second in each pair. If no schedule is provided, one will be determined automatically upon initialization.

The \texttt{color} argument specifies the bicolouring of the qubits. In the $d=3$ example, qubit $3$ should be coloured differently to qubits $0$ and $6$. This can be expressed as follows.
\begin{verbatim}
color = {0:0, 3:1, 6:0}
\end{verbatim}
Here qubits $0$ and $6$ are assigned the `color' \texttt{0} while qubit $3$ is coloured \texttt{1}. If no colouring is provided, one will be determined automatically upon initialization. For some link graphs, no bicolouring is possible in which all neighbouring code qubits are differently coloured. The bicolouring will therefore be a compromise. The argument \texttt{max\_dist}, which has a default value of $2$, is used in the calculation of this compromise, and different values can yield different results.

Further keyword arguments are as follows.
\begin{itemize}
\item \texttt{basis}: basis used for encoding. The two values specify the basis for the two colours. Default is \texttt{'xy'}.
\item \texttt{logical}: the logical value to be encoded, either \texttt{'0'} or \texttt{'1'}. Default is \texttt{'0'}.
\item \texttt{delay}: time for a delay to add after mid-circuit measurements. Default is none.
\item \texttt{barriers}: whether to include barriers between different sections of the code. Default is \texttt{True}.
\item \texttt{run\_202}: whether to run $[[2,0,2]]$ sequences. This will be overwritten if $T$ is not high
            to cover all links. Default is \texttt{True}.
\item \texttt{rounds\_per\_202}: number of rounds that are part of $[[2,0,2]]$ sequences, including the typical link
            measurements at the beginning and end. At least 9 are required to get an event dedicated to
            conjugate errors. Default is $9$.
\item \texttt{ff}: whether to correct the effects of [[2,0,2]] sequences via feedforward. Default is \texttt{True}.
\item \texttt{conditional\_reset}: whether to apply conditional resets (an \texttt{x} gate conditioned on the result of the previous measurement), rather than a standalone reset gate. Default is \texttt{False}.
\end{itemize}

Upon initialization, an \texttt{ArcCircuit} object creates the circuits for the code, which initialize the logical value, implement the syndrome measurement rounds and then perform final readout. These can be accessed with \texttt{code.circuit}, which takes the form of a dictionary with the encoding basis as keys and the Qiskit circuit objects as values. The circuit for the given encoding can then be accessed with \texttt{code.circuit[code.basis]}, and the conjugate encoding with \texttt{code.circuit[code.basis[::-1]}.

The \texttt{ArcCircuit} object also provides the means to transpile circuits to specific devices. For this, the devices need to be represented by a Qiskit backend object. For the device \texttt{ibm\_lagos} used as the basis of the $d=3$ example above, the corresponding backend object is obtained as follows.
\begin{verbatim}
from qiskit_ibm_provider import IBMProvider
provider = IBMProvider()
backend = provider.get_backend('ibm_lagos')
\end{verbatim}
The transpilation is then performed using \texttt{code.transpile(backend)}, which provides a dictionary of the transpiled circuits. By default, pairs of \texttt{x} gates are inserted into periods of delay on the code qubits for dynamical decoupling. This is not applied to the auxiliary qubits by default since, ideally, they should spend the entire circuit in state $|0\rangle$. However, additional arguments exist to allow other forms of dynamical decoupling.

A single circuit is run in Qiskit using the following.
\begin{verbatim}
job = backend.run(circuit, shots=10000)
\end{verbatim}
Here \texttt{circuit} is the circuit, and \texttt{shots} is the number of samples to use. The results can be extracted from the \texttt{job} object as a counts dictionary with \texttt{job.result().get\_counts()}. For the $d=3$ example with \texttt{T=2}, an example of a counts dictionary is as follows.
\begin{verbatim}
counts = {'000 00 00': 9679, '100 10 00': 321}
\end{verbatim}
The strings here represent, from right to left, the results of of the two syndrome measurement rounds followed by the final readout. The corresponding numbers are the number of shots for which this string results. The most common string here represents the case of no errors and a logical \texttt{0}. The other represents an error on qubit $6$ between the first and second syndrome measurement rounds. It is detected by one of the measurements in the second round, and is evident from the final readout of this code qubit.

\section{Example results} \label{sec:results}

We will now consider examples of results from ARCs run on IBM Quantum superconducting qubits. A large number of preliminary experiments were run from late 2022 to early 2023~\cite{milan-thesis}. With the experience gained from these, a realistic scope for a straightforward and realistic set of experiments was determined.

\begin{enumerate}
\item Experiments will focus on an entire device, rather than an optimally chosen subset of qubits. When a subset is required, the qubits are chosen arbitrarily.
\item Devices will be used `out of the box', with no special tuning to the needs of a device beyond what is applied by the transpiler,
\item Repetition codes and $[[2,0,2]]$ codes will be run separately.
\item Experiments will focus on 127 qubit \textit{Eagle} devices.
\item The number of syndrome measurement rounds will be $T=10$.
\end{enumerate}

Point 1 here has been chosen to keep the experiments straightforward. Finding optimal subsets of qubits could be used to increase the quality of results, but the overhead in terms of running experiments and their analysis is high. This process would also need to be repeated whenever new results are taken, to account for changes in the behaviour of the devices. Using less qubits also allows more syndrome measurement rounds to be implemented before the control systems are overwhelmed, which poses another tradeoff to be considered. All this complexity would obscure a simple truth: a large device is only a large device if all the qubits can work together effectively. We therefore focus on devices as a whole.

Point 2 also helps to decrease complexity. IBM Quantum devices support alternative forms of some gates, such as the reset of a qubit. Large numbers of resets and measurements throughout a circuit can lead to jobs failing due to problems within the control system, but experimenting with different forms of reset could mitigate this. As could the addition of delay gates after these operations, which serve to space them out further. All options come with tradeoffs, which demand overheads of experiments and analysis. For the implementation of fault-tolerant quantum computation, such analysis is clearly highly important, and will help drive forward development of the hardware. However, from the perspective of a third-party user looking to run proof-of-principle quantum error correction experiments, such in-depth analysis is likely to be out of scope. We therefore take the perspective of such a user during the implementation of ARCs presented here, using the devices `out-of-the-box' with the standard Qiskit implementations of gates.

Point 3 contradicts an important intention for ARCs, which is that they should be run with the $[[2,0,2]]$ codes interleaved into the repetition code. However, this places a large demand on the number of syndrome measurement rounds. There are at least 9 rounds required per link used for a $[[2,0,2]]$, and since \textit{Eagle} devices have 71 links, several hundred syndrome measurement rounds are required. However, the maximum number of rounds  that were successfully implemented on a whole \textit{Eagle} device in the preliminary experiments was not sufficient to implement this. It is therefore necessary to implement the two parts of the experiment separately, with repetition codes running across the whole device and $[[2,0,2]]$ codes implemented on only a subset of qubits. In accordance with Point 1, the qubits for the $[[2,0,2]]$ are arbitrarily chosen to be essentially be the first seven qubits in the device to lie along a line.

There are a wide range of different IBM Quantum systems, belonging to different revisions of different processor families. Some have distinct exploratory features. However, it is the \textit{Eagle} family that is currently the workhorse of the IBM Quantum fleet in the near-term. This is the reason for the focus on these in Point 3. Furthermore, to run $[[2,0,2]]$ codes we require devices that support OpenQASM 3 and dynamic circuits~\cite{cross-22}. At the time of taking results, only one such device was in service: \texttt{ibm\_sherbrooke}. All results presented will therefore be from this device.

The number of syndrome measurement rounds set in Point 5 is chosen both because it is a realistic goal for other hardware approaches (see \cite{sundaresan-23,moses-23}, which both use around this number), and because it is the minimum number of rounds required for the $[[2,0,2]]$ experiment.

The results presented below were from a single set of runs on arbitrary dates. There was no optimization over results taken or day-to-day device properties. They are intended to serve as a snapshot of an arbitrary point in time, as ARCs could theoretically provide on a daily basis.

\subsection{Microscopic results for repetition codes}

\begin{figure}
    \includegraphics[width=0.45\textwidth]{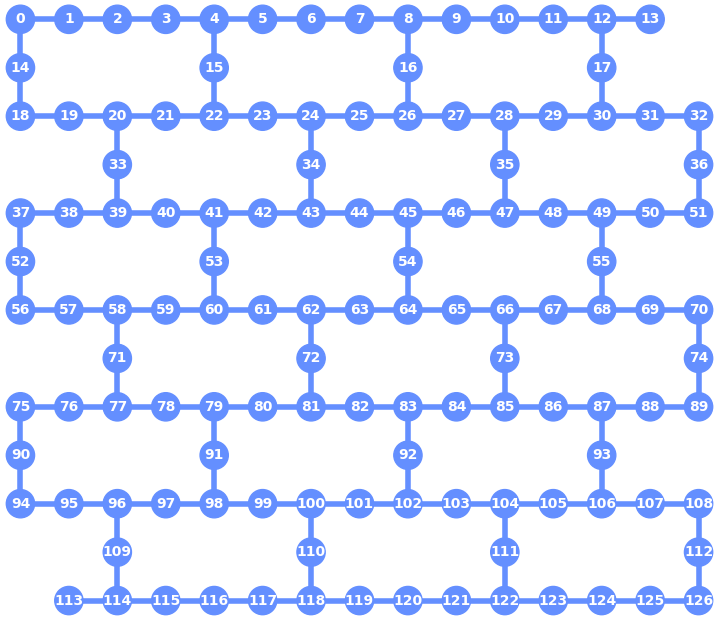}
    \caption{Layout of \texttt{ibm\_sherbrooke}, showing the numbering used for the qubits.}
    \label{fig:sherbrooke}
\end{figure}

Repetition codes were run on the \texttt{ibm\_sherbrooke} device, belonging to the \textit{Eagle} family of 127 qubit quantum processors. These devices have a layout in which qubits lie on the vertices and edges of a honeycomb lattice (see Fig. \ref{fig:sherbrooke}). This is naturally suited to a decomposition into links, with the qubits on vertices serving as code qubits, and those on edges as auxiliaries. Only two qubits are not so lucky -- those numbered 13 and 113 -- since they are situated on a truncated edge of the honeycomb lattice. This means that the repetition codes on this device use a total of 125 qubits. Of these 54 are code qubits and 71 are auxiliaries. The circuits were run for $T=10$, and both with and without resets.

The code was implemented using \texttt{basis='xz'}, which means that runs for both the $XZ$ and $ZX$ encoding were made. This was chosen so that the probabilities deduced for code qubits could be regarded as being those for bit-flip errors and phase-flip errors. Codes with both \texttt{logical='0'} and \texttt{logical='1'} were implemented. The circuits for these four possibilities were implemented in sequence, with this process then repeated for 10000 shots. Dynamical decoupling was applied only to the code qubits using the default settings. Variants of the circuit with and without resets after measurements were run. In the latter case, a delay of the same length as a reset was added. All circuits were run on 3rd July 2023.

The error probabilities for each edge were calculated for the cases both with and without resets. A representative error rate for each qubit over the entire circuit can then be obtained by averaging over all edges with support on that qubit. 

\begin{figure}
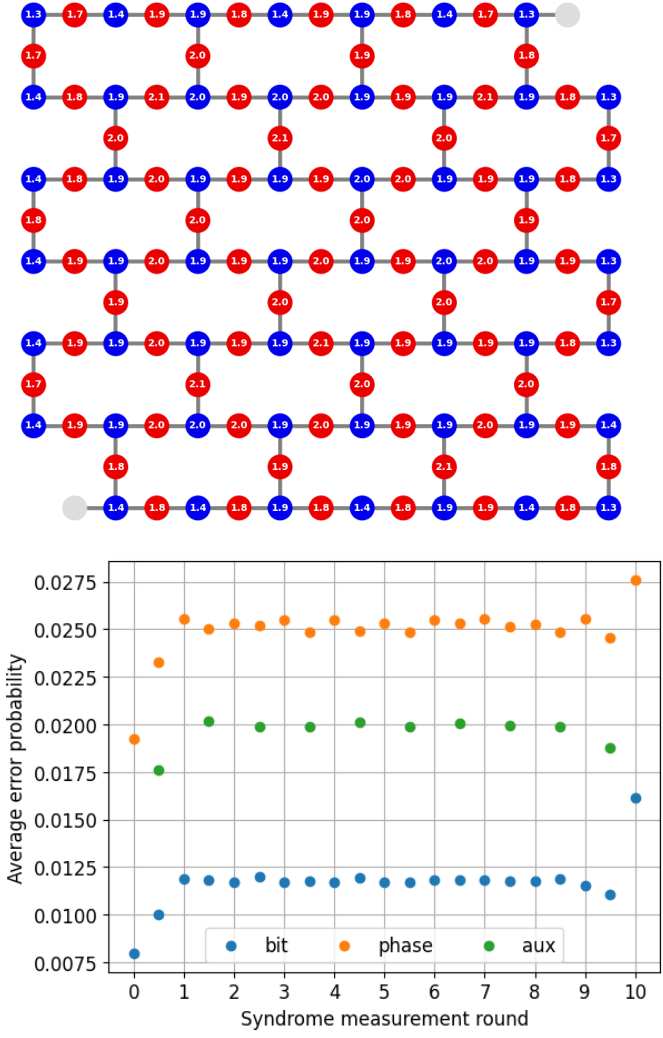

    \subfigure{\includegraphics[width=0.45\textwidth]{figs/1348f289-c591-4035-9df1-6bb5c652bac7\_None.png}}
    \subfigure{\includegraphics[width=0.5\textwidth]{figs/sim\_micro.png}}
    \caption{Results for simulated noise with $p=1\%$. (a) Probabilities of errors associated with each qubit, averaged over all rounds and both basis variants. Code qubits are shown in blue and auxiliaries in red. Colours are scaled such that errors of $25\%$ or are black, and the colour becomes progressively brighter down to $0\%$. The two unused qubits are shown in grey. Small variations in values can be attributed to statistical errors. (b) Average probabilities of errors plotted against the time period associated with the error.}
      \label{fig:sim_errors}
\end{figure}

First, let us consider the results for simulated errors shown in Fig. \ref{fig:sim_errors}. These are obtained for an error model in which all errors occur with the same probability. Specifically, single qubit depolarizing noise is applied to all single qubit gates, two qubit depolarizing noise is applied to two-qubit gates and an \texttt{x} error is applied prior to all measurements, all independently with probability $p=1\%$. The depolarizing noise contributes an error rate of $p/2$ on each qubit for both bit- and phase-flip errors.

Fig. \ref{fig:sim_errors} (a) shows the average error for each qubit, averaged over all adjacent edges, logical values and basis choices. Fig. \ref{fig:sim_errors} (b) shows the errors at different times throughout the circuit, with results for bit- and phase-flips on code qubits, as well as auxiliary qubits, plotted separately. 
For the auxiliary qubits, errors come from two sources: the two CNOTs and the measurement. The total error probability for the measurements is then $\sim 2p$. This is clearly seen in Fig. \ref{fig:sim_errors} (b).

For the code qubits there is not such a clear specific value for the error rates. This is because the local CNOT schedule, and whether a code qubit is involved in two or three CNOTs, will determine how much error can occur between different points at which distinct error signatures can be detected. Error events detected by hyperedges will also contribute to multiple edges, and so such the probability of such errors will be overcounted. However, one aspect of their behaviour that should be clear from the simulations is the difference between bit- and phase-flip errors. This is because the CNOTs are conjugated by Hadamards on qubits sensitive to phase-flip noise, and so each will contribute an error rate of $\sim 1.5 p$ rather than just $p/2$. Error rates for phase-flips should therefore be significantly higher than those for bit flips in the simulatuons. This is clearly seen in Fig. \ref{fig:sim_errors} (b).

\begin{figure}
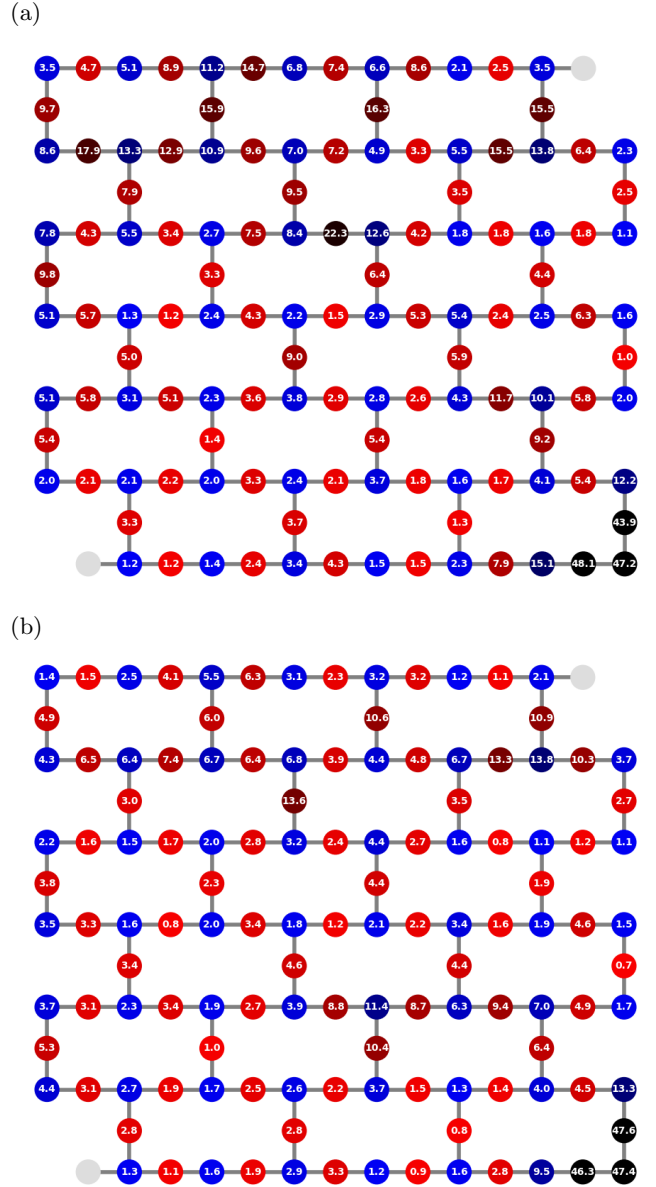

    \begin{flushleft}(a)\end{flushleft}
    \subfigure{\includegraphics[width=0.45\textwidth]{figs/ciff837985671v40uq40\_None.png}}
    \begin{flushleft}(b)\end{flushleft}
    \subfigure{\includegraphics[width=0.45\textwidth]{figs/cifg0vc60l54kv9qhi30\_None.png}}
    \caption{Probabilities of errors associated with each qubit, averaged over all rounds and both basis variants. Code qubits are shown in blue and auxiliaries in red. Colours are scaled such that errors of $25\%$ or are black, and the colour becomes progressively brighter down to $0\%$. The two unused qubits are shown in grey. Results are for (a) \texttt{resets=True} and (b) \texttt{resets=False}.}
      \label{fig:device_errors}
\end{figure}

Results are shown for data from \texttt{ibm\_sherbrooke} in Fig. \ref{fig:device_errors}, for runs both with and without post-measurement resets. In both cases, the results demonstrate that the majority of the device works well, with average error rates of a few percent. However, it is evident that certain regions, particularly around qubit 126, have significant errors during the process.

\begin{figure}
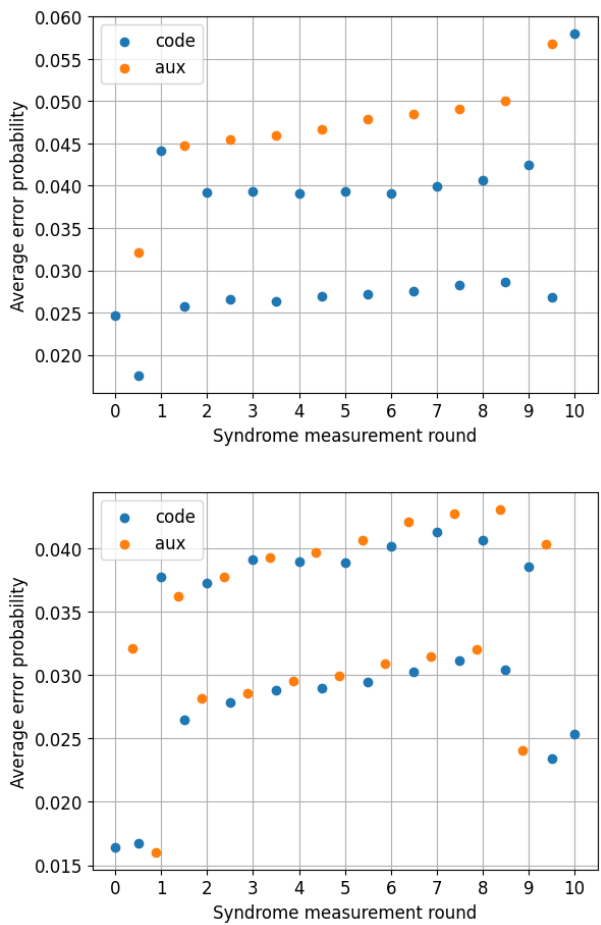

    \subfigure{\includegraphics[width=0.45\textwidth]{figs/ciff837985671v40uq40\_micro.png}}
    \subfigure{\includegraphics[width=0.45\textwidth]{figs/cifg0vc60l54kv9qhi30\_micro.png}}
    \caption{Average probabilities of errors for (a) \texttt{resets=True} and (b) \texttt{resets=False}, plotted against the time period associated with the error.}
      \label{fig:micro_time}
\end{figure}

For the most well-behaved qubits, i.e. those for which the average is less than $10\%$, we also consider the average error probabilities over time. Specifically, we average over all qubits for each point in time, as depicted in Fig. \ref{fig:micro_time}. For the code qubits, the time for each type of error is rounded to each round or half round.

In Fig. \ref{fig:micro_time} (a), the errors for auxiliary qubits are plotted halfway between rounds. This is to reflect the fact that these errors can occur throughout the round: either on the auxiliary qubit before the measurement or during the measurement. In Fig. \ref{fig:micro_time} (b) there are two different sources of measurement error that can be distinguished. The ones shown earlier in the round depict errors that flip the auxiliary qubit state either before or during the measurement. The ones shown toward the end of the round depict errors due to misassignment of the readout outcome. From these results we see that the majority of measurement errors are due to flips of the auxiliary qubit state, rather than misassignment.

The treatment of errors on code qubits is the same for both the case with and without resets. The errors shown at the beginning of each round depict those that occur between the CNOT gates of two measurement rounds, including those that occur during the measurement. These are primarily idling errors, and so the probability is understandably larger. Those shown at the middle of the round correspond to errors that occur between the CNOT gates of the same measurement round, and hence are detected in different measurement rounds. These are primarily CNOT errors or idling during the shorter times between CNOTS, and so are understandably lower.

In the cases of both auxiliary and code qubit errors, an increase in error rates over time is seen. For auxiliary errors, this may be due to the increasing probability of leakage. For the code qubit errors, it is difficult to tell if the apparent increase induced by the increase of associated measurement errors, or if the code qubit error rate is actually increasing.

The error rates at the beginning and end of the process are seen to be much different than those during the bulk of the circuit. This is due to the initial errors being calculated from syndrome changes involving only one measurement result, and so being less affected by measurement errors. Also, the final syndrome round has the syndrome inferred by direct measurement of the code qubits, meaning that the syndrome changes involve three measurement results, and so can be more affected by measurement errors.

\subsection{Macroscopic results for repetition codes} \label{subsec:macro}

The macroscopic analysis was applied to the device-wide $T=10$ results from \texttt{ibm\_sherbrooke}, as described earlier. This analysis was done by means of the Union Find decoding algorithm all four instances of the code (with each basis, and each logical value). To provide context for results from \texttt{ibm\_sherbrooke}, we will first consider simulated results for a code defined on the \textit{Eagle} architecture with $T=10$, and with the simple error model described previously.

\subsubsection{Results from simulation}

\begin{figure}[h]
    \begin{flushleft}(a)\end{flushleft}
    \subfigure{\includegraphics[width=0.45\textwidth]{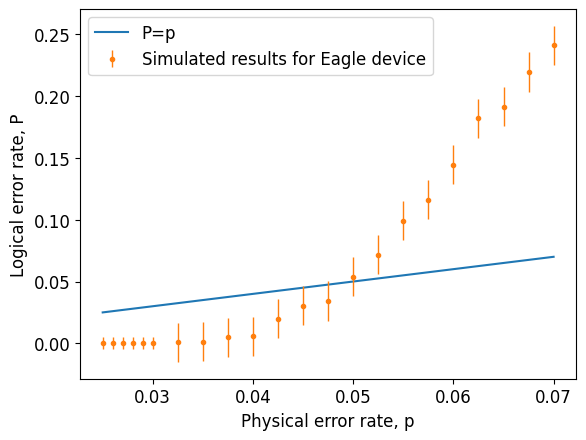}}
    \begin{flushleft}(b)\end{flushleft}
    \subfigure{\includegraphics[width=0.45\textwidth]{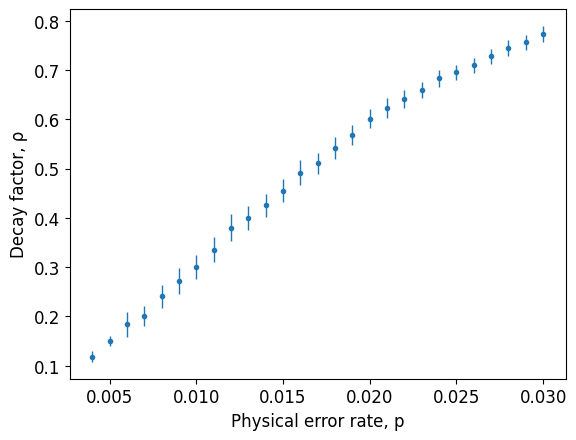}}
    \caption{For simulated results on an ARC circuit defined for an \textit{Eagle} device, (a) the logical error versus physical error rate, and (b) the decay rate versus physical error rate.}
    \label{fig:sim_log}
\end{figure}

 Results from simulated results are shown in Fig. \ref{fig:sim_log}. Since the error model we consider has a single error probability $p$, we can define the pseudothreshold for the code as the point at which the $p$ is equal to the logical error rate of the code. From Fig. \ref{fig:sim_log} (a) it can be seen that this occurs at around $p=5\%$. The decay factors below threshold, obtained by fitting an exponential decay to the decay of cluster frequency with error number, are shown in \ref{fig:sim_log} (b). Well below threshold we see that $\rho \approx 30p$, confirming the expected linear relationship.

\begin{figure}[h]
    \begin{flushleft}(a)\end{flushleft}
    \subfigure{\includegraphics[width=0.45\textwidth]{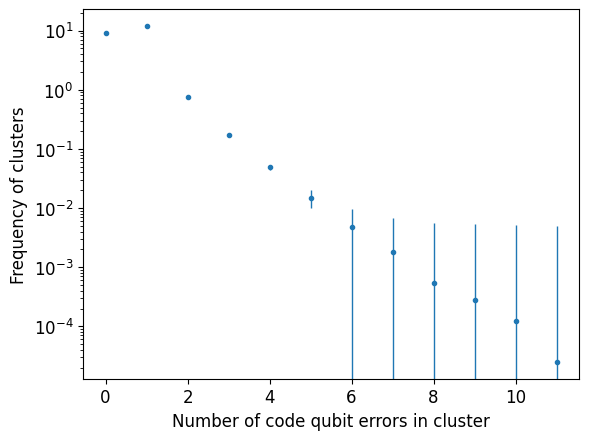}}
    \begin{flushleft}(b)\end{flushleft}
    \subfigure{\includegraphics[width=0.45\textwidth]{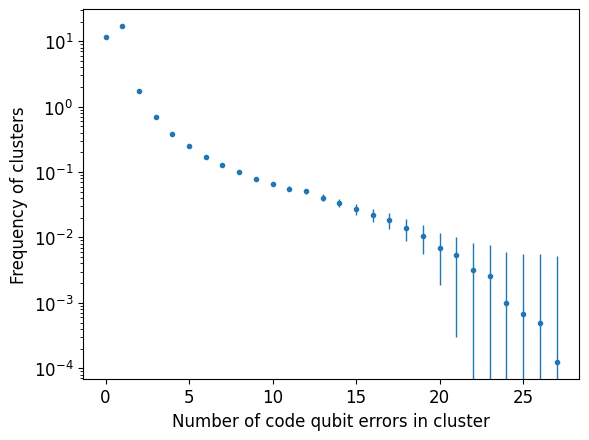}}
    \caption{Cluster frequencies versus error number for simulated results on an ARC circuit defined for an \textit{Eagle} device with (a) $p=1\%$, and (b) $p=3\%$. Error bars depict $1/\sqrt{N}$, where $N$ is number of samples.}
    \label{fig:sim_nums}
\end{figure}

The relationship between cluster frequencies and cluster size are shown in Fig. \ref{fig:sim_nums}. Here the frequency is the number of clusters of a given size found over all samples, divided by the number of samples. The size of a cluster is defined as the minimum number of code qubit errors required to create the flattened form of a cluster.

For the case of $p=1\%$, the results are clearly consistent with an exponential decay. To determine $\rho$, a linear fit is made onto the logarithms of the frequencies, yielding a gradient of $\ln \rho = -1.20 \pm 0.04$, and hence $\rho=0.3$. Such a clear decay is typical of results for $p \leq 2\%$.

For the case of $p=3\%$, a clear decay is also seen. However, it is less obviously exponential. Instead there is a hump in the middle, likely due to the increased relevance of entropic factors causing it to become more likely that mid-sized clusters will merge to form larger clusters. This hump becomes ever more apparent for $p>2\%$, making it harder to accurately determine the decay rate with the given data and leading to the kink of Fig. \ref{fig:sim_log} (b).

\subsubsection{Results from \texttt{ibm\_sherbrooke}}

ARC circuits were implemented for $T=10$ on \texttt{ibm\_sherbrooke}, both with and without resets.

In each case decoding was run on all four instances of the code (with each basis, and each logical value). No logical errors were found for all $10^4$ samples of each, allowing us to confidently use the clustering provided by the decoder to analyze the errors.

\begin{figure}[h]
    \begin{flushleft}(a)\end{flushleft}
    \subfigure{\includegraphics[width=0.45\textwidth]{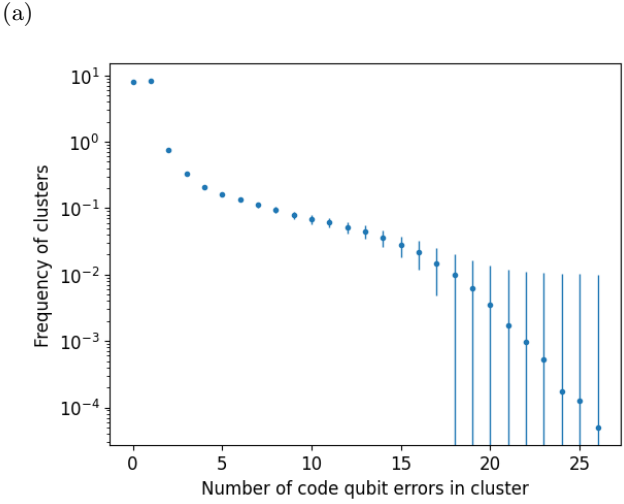}}
    \begin{flushleft}(b)\end{flushleft}
    \subfigure{\includegraphics[width=0.45\textwidth]{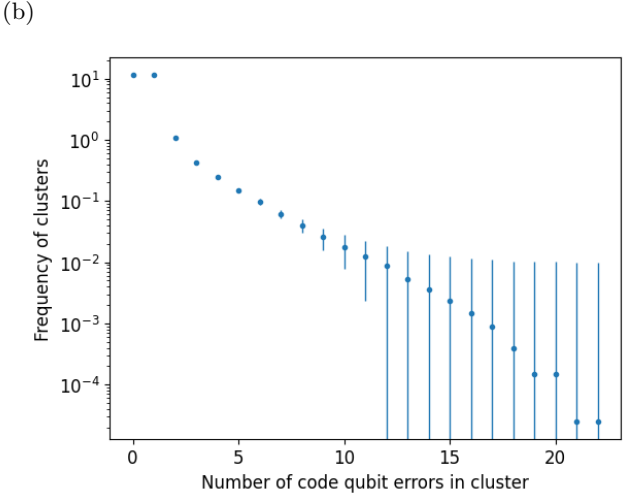}}
    \caption{Cluster frequencies versus error number for results from  \texttt{ibm\_sherbrooke} with $T=10$ and (a) \texttt{resets=True}, and (b) \texttt{resets=False}. Error bars depict $1/\sqrt{N}$, where $N$ is number of samples.}
    \label{fig:nums10}
\end{figure}

The size of the clusters found by the decoder are shown in Fig. \ref{fig:nums10} (a) for the case with resets. Here we see an initial decay of cluster size, followed by a hump, and then a further decay. It is notable that, at the very tail of this decay, a handful of shots are found with almost $d/2$ errors. Fitting to an exponential decay yields $\ln \rho = -0.252 \pm 0.014$, and so $\rho=0.78$. Both the decay curve and the value of $\rho$ are quite similar to those of the simulated data for $p=3\%$.

Results for the case of no resets are shown in Fig. \ref{fig:nums10}(b). These exhibit a much clearer exponential decay, with  $\ln \rho = -0.44 \pm 0.01$ and $\rho=0.64$. This value is similar to that for simulations with $p=2.3\%$.

\begin{figure}
    \includegraphics[width=0.45\textwidth]{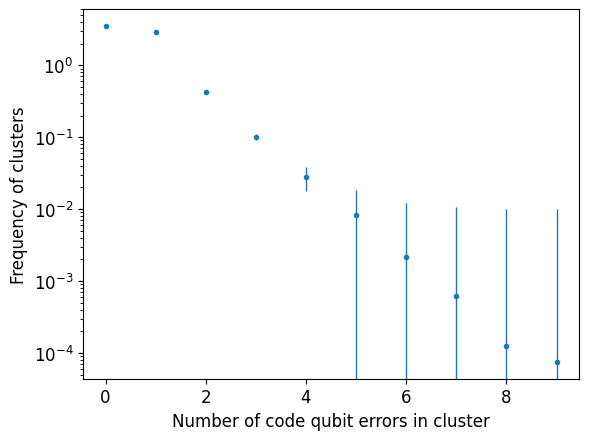}
    \caption{Cluster frequencies versus error number for results from  \texttt{ibm\_sherbrooke} with $T=1$ and (a) \texttt{resets=True}, and (b) \texttt{resets=False}. Error bars depict $1/\sqrt{N}$, where $N$ is number of samples.}
    \label{fig:nums1}
\end{figure}

Results were also obtained from \texttt{ibm\_sherbrooke} for the case of $T=1$. These are shown in Fig. \ref{fig:nums1}. The fit yields $\ln \rho = -1.31 \pm 0.02$ and $\rho=0.27$.

\subsection{Results for $[[2,0,2]]$s}

For a minimal implementation of $[[2,0,2]]$ codes we require three links in a line, with the central one alternating between the two syndrome measurement types and the others used to define the classically controlled feedforward operation which `cleans up' after the process. A minimal of 9 rounds are required in all, but we instead use \texttt{T=10} so that errors in the final readout can be more easily distinguished from those due to faulty feedforward operation.

The code was implemented using \texttt{basis='xy'}, which means that code qubits are primarily in a superposition state and will be vulnerable to both relaxation and dephasing errors. As such, runs for \texttt{logical='1'} were not deemed necessary. Dynamical decoupling was applied to all qubits with a pair of \texttt{x} gates inserted into each delay. The jobs were run on \texttt{ibm\_sherbrooke} on 5th June 2023. The following \texttt{links} and \texttt{schedule} were used:
\begin{eqnarray} \nonumber
&&\mathtt{[(0,1,4),(4,7,10),(10,12,15)] }, \\ \nonumber
&&\mathtt{[[(0,1),(4,7),(10,12)],[(4,1),(10,7),(15,12)]]}.
\end{eqnarray}

The probabilities for the two edges that could correspond to feedforward errors were $18.1\%$ and $15.3\%$, both with a standard error of $0.4\%$. The probability of the conjugate error was $12.6 \pm 0.4\%$, compared to the two corresponding standard errors probabilities which were both $11.1 \pm 0.4 \%$. From these numbers alone we see that the conjugate error probability agrees well with those for standard errors, being only slightly greater. The feedfoward errors however, are markedly higher.

\begin{figure}
    \includegraphics[width=0.45\textwidth]{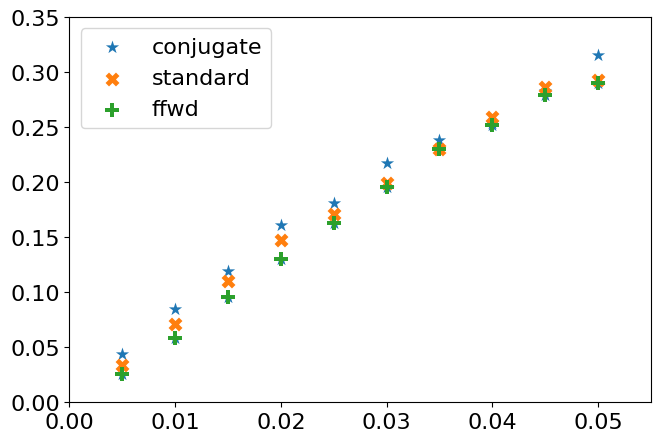}
    \caption{Probability of various edge error probabilities for the $[[2,0,2]]$ codes plotted against the error rate, $p$, of a simple error model.}
    \label{fig:202_graph}
\end{figure}

To further understand these results, a baseline is required. For this, simulations were performed with the simple error model described previously. Specifically, single qubit depolarizing noise was applied to all single qubit gates, two qubit depolarizing noise was applied to two-qubit gates and an \texttt{x} gate was applied prior to all measurements, all independently with probability $p$. The edge error probabilities were then averaged between the \texttt{'xy'} and \texttt{'yx'} basis versions. The results for the most pertinent edges - those for the conjugate errors, the comparable standard errors and those resulting from faulty feedforward -- are shown in Fig. \ref{fig:202_graph}.

We see that the $\sim 10\%$ probabilities of standard and conjugate errors in the \texttt{ibm\_sherbrooke} results are similar to those of $p \approx 1.5\%$, whereas the feedfoward probabilities are more similar to those of $p \approx 2.5\%$. We therefore find that the system behaves in a manner consistent with an error rate of a few percent, even while implementing classically controlled operations and simultaneously detecting a full set of errors.

\section{Outlook}

In this work we propose repetition codes with a range of enhancements, and implement them with 125 qubits, a distance of 54 and with 10 syndrome measurement rounds. The scale of these experiments are such that similar implementations could be achieved with other current and near-term quantum hardware, allowing cross-platform comparisons.

Future work should also aspire to increase the scale of the experiments. In particular, an implementation in which enough rounds are implemented to apply the $[[2,0,2]]$ process on all links would demonstrate an impressive degree of control over the resources required for fault-tolerance.

\section{Acknowledgements}

The authors would like to thank Drew Vandeth, Andrew Cross, Grace Harper and other contributors to Qiskit QEC, as well as Edward Chen, Riddhi S. Gupta, Bence Het\'enyi, Luke Govia, David McKay, Seth Merkel and others at IBM Quantum for comments throughout the development of this work.

Research was sponsored by the Army Research Office and was accomplished under Grant Number W911NF-21-1-0002. The views and conclusions contained in this document are those of the authors and should not be interpreted as representing the official policies, either expressed or implied, of the Army Research Office or the U.S. Government. The U.S. Government is authorized to reproduce and distribute reprints for Government purposes notwithstanding any copyright notation herein.

\section{Data and Code Availability}

The most up-to date version of the software in this work can be found within Qiskit QEC~\cite{qiskitqec}. The notebooks used to collect data, and the data itself, can be found at \cite{source}.

\bibliography{references}

\end{document}